\documentclass[12pt]{article}
\usepackage{mathrsfs}
\usepackage{amsmath}
\usepackage{mathrsfs}
\usepackage{amssymb}
\usepackage{color}
\usepackage{psfrag}
\usepackage{graphicx}
\usepackage{subfigure}
\usepackage{rotating}
\usepackage{cite}
\usepackage[colorlinks=true,linktocpage=true,linkcolor=blue,citecolor=blue]{hyperref}
\usepackage[section]{placeins}
\usepackage{amsfonts}
\usepackage{ifpdf}
\usepackage[toc,page]{appendix}
\ifpdf
\else

\newcommand{\bea}{\begin{eqnarray}}
\newcommand{\eea}{\end{eqnarray}}
\newcommand{\be}{\begin{equation}}
\newcommand{\ee}{\end{equation}}
\renewcommand{\a}{\alpha}
\renewcommand{\b}{\beta}
\newcommand{\g}{\gamma}

\newcommand{\de}{\delta}

\newcommand{\la}{\lambda}

\newcommand{\ka}{\kappa}
\newcommand{\om}{\omega}
\newcommand{\Om}{\Omega}
\newcommand{\half}{\frac{1}{2}}
\newcommand{\pa}{\partial}
\newcommand{\si}{\sigma}
\newcommand{\Si}{\Sigma}
\renewcommand{\t}{\theta}
\newcommand{\nn}{\nonumber}
\newcommand{\lan}{\langle}
\newcommand{\ran}{\rangle}
\newcommand{\z}{\mathcal}

\newcommand{\go}{\gamma_{0}^{(i)}}
\newcommand{\ga}{\gamma^{(i)}}
\newcommand{\ko}{\kappa_{0}^{(i)}}
\newcommand{\Do}{D^{(i)}}
\newcommand{\mo}{\mu^{(i)}}

\newcommand{\vs}[1]{\vspace{#1 mm}}

\begin{document}
\topmargin 0pt
\oddsidemargin 0mm

\vspace{2mm}

\begin{center}

{\Large \bf {Brownian motion in strongly coupled, anisotropic Yang-Mills plasma: A holographic approach}}

\vs{10}

{Shankhadeep Chakrabortty$^{a}$\footnote{E-mail: shankha@imsc.res.in}, Somdeb Chakraborty$^{b}$\footnote{E-mail: somdeb.chakraborty@saha.ac.in}and Najmul Haque$^{b}$\footnote{E-mail: najmul.haque@saha.ac.in}}

 \vspace{4mm}

{\em
$^{a}$Institute of Mathematical Sciences, IV Cross Road, CIT Campus, Taramani, Chennai-600113, India\\
\vs{4}
 $^{b}$Saha Institute of Nuclear Physics,  1/AF Bidhannagar, Kolkata-700 064, India\\}

\end{center}

\vs{10}

\begin{abstract}
We employ methods of gauge/string duality to analyze the non-relativistic Brownian motion and the concomitant Langevin equation of a heavy quark
 in a strongly coupled, thermal,  anisotropic Yang-Mills plasma in the low anisotropy limit. We consider fluctuations both along and perpendicular to the direction 
of anisotropy and study the effects of anisotropy on the drag coefficient, the diffusion constant and the Langevin coefficient for both 
the directions. We also verify the fluctuation-dissipation theorem for Brownian motion in an anisotropic medium. 
\end{abstract}

\newpage
\tableofcontents
\section{Introduction}
A particle immersed in a hot fluid exhibits an incessant, random dynamics known as Brownian motion \cite{Brown}.
The Brownian motion originates from the collisions experienced by the particle with the constituents of 
the fluid undergoing a  random thermal motion. 
The consideration of these random collisions requires the fact that the fluid medium is not a continuum but made of finite-size constituents. Hence, the Brownian motion
actually offers a better understanding of the underlying microscopic physics of the medium. The random dynamics of a Brownian particle
is encoded in the Langevin equation describing the total force acting on the particle as a sum of dissipative and random forces. Although both 
of these forces have the same microscopic origin, phenomenologically the dissipative
force describes the in-medium frictional effect and the random force stands for a source of random kicks from the medium.

Brownian motion is a universal phenomenon for all finite temperature systems. Therefore, a heavy probe quark immersed in a strongly coupled hot quark-gluon plasma (QGP) which is believed to be created in the RHIC and LHC experiments \cite{RHIC},
 undergoes the same thermal motion \cite{Rapp:2009my}. From field theoretical standpoint, the random motion 
in QGP phase is hard to study due to non-perturbative strong coupling effects. 
However, the $AdS/CFT$ correspondence \cite{Maldacena:1997re, Witten:1998qj, Gubser:1998bc, Aharony:1999ti} 
seems to be a good theoretical tool in this regard, since it has been extensively used to 
study a large class of strongly coupled plasma having well-defined gravity duals. 
In spite of intensive efforts, till  date, the gravity dual of strongly coupled QGP phase remains
elusive and the gauge theories having well defined gravity duals are different
from QGP in several aspects. Nonetheless, it is remarkably 
found in some instances that, many strong coupling features extracted holographically from known geometric duals for UV 
conformal theories agree with the thermal QGP phase. For example, in \cite{1,2}, the $AdS/CFT$ correspondence has been used 
to show that the shear viscosity to the entropy
ratio of four-dimensional $SU(N_{c})$ ($N_{c}$ being the number of colors)  Yang-Mills theory with $\mathcal{N} = 4$ supersymmetries is $1/4\pi$.
This low viscosity is also speculated from the estimation of the RHIC data for QGP \cite{Teaney:2003kp}. 
Later the ratio was found to be universal for all the strongly coupled gauge 
theories, in the $N_{c} \rightarrow \infty$ limit, having a gravity dual \cite{3}. Subsequently, it was found that
there are other physical quantities, such as, $R$-charge conductivity to charge susceptibility ratio, a certain combination of thermal conductivity,
temperature and chemical potential, that
show universal behavior too \cite{3,4}. Motivated by these universal outcomes, there has been a substantial amount of 
holographic analysis of dissipative physics of various thermal plasma having dual gravity to understand the dynamical 
feature of QGP phase in a better way, see, for example, \cite{Herzog:2006gh, Gubser:2006bz, Caceres:2006dj, Matsuo:2006ws, Chernicoff:2012iq, Herzog:2007kh, VazquezPoritz:2008nw, Sadeghi:2009mp, Roy:2009sw, Panigrahi:2010cm, Chakrabortty:2011sp, Cai:2012eh, CasalderreySolana:2006rq, Gubser:2006nz, CasalderreySolana:2007qw,Chernicoff:2012gu,Chernicoff:2012bu,Chakraborty:2012dt,Chakraborty:2012pc,Giataganas:2012zy,Giataganas:2013lga}. Recently, as an important improvement in this direction, the Brownian motion of a probe particle has been successfully studied using the framework of the $AdS/CFT$ correspondence\cite{deBoer:2008gu, Son:2009vu}.

The bulk interpretation of the Brownian motion of a heavy probe quark immersed in a $SU(N_{c})$ Yang-Mills theory with 
$\mathcal{N} = 4$ supersymmetries emerges from the consideration of a probe fundamental string in the dual $AdS$ black hole background,
stretching between the $AdS$ boundary and the horizon. The end point of the string attached to the boundary is holographically mapped to the
boundary probe quark. The transverse modes of the probe string are thermally excited by the black hole environment.
This excitation propagates up to boundary and holographically incorporates the Brownian motion of boundary quark. 
In an intuitive way, the fact that, semiclassically, the traverse string modes are thermally excited by Hawking radiation reflects the 
bulk interpretation of random force in the boundary Langevin equation. On the other hand, the fact that the string excitation is absorbed 
by the black hole environment stands for the bulk realization of boundary frictional force. 
In the detailed course of computation, we need to quantize the transverse string modes. As explained in \cite{Lawrence:1993sg}, 
the Hawking radiation associated with the string excitations occurs upon quantizing these modes. 
Once these modes are quantized, using holographic prescription, 
the erratic motion of string end point attached to the boundary can be realized as the Brownian motion.

There are two independent approaches available in the literature to obtain
these results. In the first approach, the state of the quantized scalar fields are identified with the 
Hartle-Hawking vacuum representing the black hole at thermal equilibrium \cite{deBoer:2008gu}. 
In the second approach, the GKPW prescription \cite {Witten:1998qj, Gubser:1998bc} of computing retarded Green's function is utilized. 
The computation of Langevin equation is done by exploring the correspondence 
between Kruskal extension of the $AdS$ black hole geometry and the Schwinger-Keldysh 
formalism \cite{Son:2009vu}. The detailed comparison between the two independent approaches is given in \cite{Hubeny:2010ry}. 
There are further generalizations
in this direction. Holographic Brownian motion has been studied in the case of charged plasma \cite{Atmaja:2010uu}, 
rotating plasma \cite{Atmaja:2012jg, Atmaja:2013gxa, Sadeghi:2013lka}, 
non-Abelian super Yang-Mills (SYM) plasma \cite{Fischler:2012ff}, non-conformal plasma \cite{Gursoy:2010aa} and $(1+1)$-dimensional strongly
coupled CFT at finite temperature \cite{Banerjee:2013rca}. 
It has  also been studied in the low temperature domian (near criticality) \cite{Tong:2012nf, Edalati:2012tc}.
The relativistic formulation of holographic Langevin dynamics has been successfully addressed in \cite{Giecold:2009cg}. Moreover, some important universality related issues regarding the Langevin
coefficients computed along the longitudinal as well as the transverse
directions to the probe quark's motion has been studied in
\cite{Giataganas:2013hwa}.

In our paper, we study the holographic Brownian motion of a heavy probe quark moving in a strongly coupled \textit{anisotropic} plasma
at finite temperature. For simplicity, we only consider the non-relativistic limit, i.e., we take $v \ll 1$ where $v$ is the velocity of the heavy quark that undergoes Brownian motion. We also take the medium to have small anisotropy and consider only the low-lying modes of the string fluctuations. These conditions are imposed only to facilitate analytical computation. The anisotropic thermal plasma we are interested in is a spatially deformed four-dimensional $\mathcal{N} = 4$ $SU(N_{c})$ 
SYM plasma at finite temperature \cite{Mateos:2011ix, Mateos:2011tv}. 
The deformation in the gauge theory has been achieved by adding a topological Yang-Mills coupling where 
the coupling parameter has a functional dependence on one of 
the three spatial boundary coordinates signifying the anisotropic direction. 
The dual bulk geometry develops an anisotropic black hole horizon and behaves as a regular solution embedded in type IIB string theory. 
The motivation for studying the Brownian motion in the context of anisotropic $\mathcal{N} = 4$ SYM  plasma
comes from experimental observations at the RHIC signifying the possible existence of a locally anisotropic phase of QGP at 
thermal equilibrium. In the heavy ion collisions, right after the plasma is formed, it
is anisotropic and also far away from equilibrium for a time $t < \tau_{out}$ .
Further, in the
temporal window $\tau_{out} < t < \tau_{iso}$ it settles down into an equilibrium state
but still does
not achieve isotropy. Thus, if one wishes to probe the early time dynamics
of the plasma
it is essential to take into consideration this intrinsic anisotropy.  In
the regime $\tau_{out} < t < \tau_{iso}$ the plasma has a significant momentum anisotropy
that leads
to an unequal expansion of the plasma in the beam direction and the
transverse directions. Although the anisotropic plasma we are interested in does not incorporate the dynamical anisotropy 
as in QGP, however
it can be a good toy model since it has a well-defined gravity dual.

With this gravity background, following \cite{deBoer:2008gu}, 
we study the bulk interpretation of the boundary Brownian motion. In particular, we explicitly compute the friction coefficient, 
the diffusion constant and the random force correlator from a holographic 
perspective when the thermal background has an inherent anisotropy and verify the 
fluctuation-dissipation theorem and the Einstein-Sutherland relation. In our bulk analysis, we include fluctuations of the probe string modes 
along both isotropic as well as anisotropic directions. We systematically study the effect of anisotropy in the low frequency limit of the thermal
fluctuation.

The paper is organized as follows.  In section \ref{boundary} we briefly review the field theoretic aspects of 
Brownian motion and follow it, in section \ref{holo}, with the holographic description of Brownian motion in anisotropic medium. 
Section \ref{holo} is divided into four subsections. In \ref{dual} we describe the gauge theory and its supergravity dual that we
 are interested in. In \ref{bulk} we discuss some generic features of the holographic formulation of the problem. In  \ref{ani} 
we perform the holographic computation for the anisotropic direction and from there the computation for the isotropic direction 
follows in a special limit which is discussed in \ref{iso}. Finally, we conclude with a discussion of our results in section \ref{con}.

\section{Brownian motion in the boundary}\label{boundary}
We begin by presenting a brief review of the field theoretic aspect of the problem following \cite{deBoer:2008gu,Fischler:2012ff,Atmaja:2010uu}. 
The simplest phenomenological model which attempts to explain the Brownian motion of a nonrelativistic particle of mass $m$ immersed in a thermal 
bath is given by the Langevin equation along the $i$-th spatial direction\footnote{We shall explicitly keep track of the direction index $i$ 
in our discussion since we need to distinguish between the anisotropic direction and the directions transverse to it.},
\be \label{langevin}
\dot{p}_{i}(t)=-\g_{o}^{(i)} p_{i}(t)+R_{i}(t),
\ee
where $p_{i}(t)=m\dot{x_{i}}$ is the nonrelativistic momentum of the Brownian particle along the $i$-th direction. 
The model, though simple, is capable of capturing the salient features of a particle undergoing Brownian motion. 
The particle is acted upon by a random force $R_{i}(t)$ arising out of its interaction with the thermal bath and, 
at the same time, it is suffering energy dissipation due to the presence of the frictional term with $\g_{0}^{(i)}$ 
being the friction coefficient. Under the effect of these two competing forces the particle undergoes random thermal motion.  
The interaction between the Brownian particle and the fluid particles at a temperature $T$ allows for an exchange of 
energy between the Brownian particle and the fluid leading to the establishment of a thermal equilibrium. In an isotropic medium the friction coefficient does not depend upon the particular space direction under consideration. 
However, if  the medium in which the particle is immersed has an anisotropy then we expect the drag coefficient along the 
anisotropic direction $\g_{0}^{||}$ to be different from that in the isotropic plane $\g_{0}^{\perp}$.

The random force $R_{i}(t)$ can be approximated by a sequence of independent impulses, each of random sign and magnitude, such that the average vanishes. Each such impulse is an independent random event, i.e., $R_{i}(t)$ is independent of $R_{i}(t')$ for $t \neq t'$. Such a noise source goes by the name of white noise. These considerations imply
\be 
\lan R_{i}(t)\ran=0, \qquad \lan R_{i}(t) R_{j}(t') \ran=\ka_{0}^{(i)}\de_{ij}\de(t-t')
\ee
where we call $\ka_{0}^{(i)}$ the Langevin coefficient. Again, the presence of anisotropy inflicts a directional dependence upon $\ka_{0}^{(i)}$. Note that, in particular, 
the random forces at two different instants are not correlated. The two parameters $\go$ and $\ko$ completely characterize the 
Langevin equation (Eq.\ref{langevin}).  As we shall see, $\g_{0}^{(i)}$ and $\ka_{0}^{(i)}$ are not independent, which is not 
unexpected since they are related by the fluctuation-dissipation theorem\footnote{The relation between the two quantities has its
 root in the fact that both the frictional force and the random force have the same origin - microscopically, they arise due to 
the interaction of the particle with the thermal medium. In this sense, the separation of the R.H.S. of Eq.(\ref{langevin}) 
in two parts is \textit{ad hoc} from the microscopic point of view, being only dictated by considerations of phenomenological simplicity.},
\be \label{flucdiss}
\g_{0}^{(i)}=\frac{\ka_{0}^{(i)}}{2mT}.
\ee
Assuming the theorem of equipartition of energy which states that each degree of freedom contributes 
$\half T$ to the energy ($T$ being the temperature and we have set the Boltzmann constant $k_{B}=1$), 
it is possible to derive the the temporal variation of the displacement squared of the particle \cite{deBoer:2008gu}
\be 
\lan s^{i}(t)^{2}\ran=\lan (x^{i}(t)-x^{i}(0))^{2}\ran=\frac{2\Do}{\g_{0}^{(i)}}\left(\g_{0}^{(i)}t-1+e^{-\g_{0}^{(i)}t} \right)
\ee
where $\Do$ is defined to be the diffusion constant. It is related to the friction coefficient $\g_{0}^{(i)}$ through 
the Einstein-Sutherland relation,
\be \label{einsuth}
\Do=\frac{T}{\g_{0}^{(i)}m}.
\ee
The solution to Eq.(\ref{langevin}) has a homogeneous part determined by the initial conditions and an inhomogeneous part proportional to the random force. The homogeneous part will decay to zero in a time of order $t^{(i)}_{\text{relax}}=1/\go$ and the long-time dynamics will be governed entirely by the inhomogeneous part, independent of the initial conditions. Based on these considerations,
one can distinguish between two different temporal domains: $t \ll 1/\go$ whence $s^{i} \sim \sqrt{T/m}\ t$ 
showing that the particle moves under inertia as if no force is acting upon it. The speed in this case is fixed by the equipartition theorem. 
In the opposite regime $t \gg 1/\go$ one obtains $s^{i} \sim \sqrt{2\Do t}$ which is reminiscent of the random walk problem. 
In this time domain, the Brownian particle loses its memory of the initial value of the velocity. The transition from one regime to another occurs at the critical value of
\be
t^{(i)}_{\text{relax}} \sim \frac{1}{\go}
\ee
which represents a characteristic time-scale of the theory, called the relaxation time, beyond which the system thermalizes.\\
The model we have considered above is based on two assumptions: i) the friction to be instantaneous and 
ii) the random forces at two different instants to be uncorrelated. The validity of these assumptions holds good only when 
the Brownian particle is very heavy compared to the constituents of the medium. However, this does not give the correct picture 
when the Brownian particle and the constituents of the medium have comparable masses. 
To overcome these pitfalls the Langevin equation is generalized such that the friction now depends upon the past history of
 the particles and also the random forces at different instants are correlated. To incorporate these effects we modify
 Eq.(\ref{langevin}) to the generalized Langevin equation,
\be \label{langgen}
\dot{p_{i}}(t)=-\int_{-\infty}^{t}dt' \g^{(i)}(t-t')p_{i}(t')+R_{i}(t)+K_{i}(t).
\ee
Note that now the history of the particle is encoded in the function $\ga(t-t')$ and we have also included the possibility of an 
external force impressed upon the particle through the term $K_{i}(t)$. $R_{i}(t)$ now obeys,
\be \label{random}
\lan R_{i}(t)\ran=0, \qquad \lan R_{i}(t) R_{i}(t') \ran=\ka^{(i)}(t-t').
\ee
At this stage it is convenient to go over to the Fourier space representation of the generalized Langevin equation
\be  \label{lanf}
p_{i}(\om)=\frac{R_{i}(\om)+K_{i}(\om)}{-i\om+\ga[\om]}
\ee
where $p_{i}(\om),R_{i}(\om)$ and $K_{i}(\om)$ are the Fourier transforms of $p_{i}(t),R_{i}(t)$ and $K_{i}(t)$ respectively, i.e.,
\be 
p_{i}(\om)=\int_{-\infty}^{\infty}dt \ p_{i}(t)e^{i\om t}
\ee
and so on. On the other hand, causality restricts $\ga(t)=0$ for $t<0$ so that $\ga[\om]$ is the Fourier-Laplace transform
\be 
\ga[\om]=\int_{0}^{\infty}dt \ga(t)e^{i\om t}.
\ee
Upon taking statistical average in Eq.(\ref{lanf}), one finds,
\be \label{adm}
\lan p_{i}(\om) \ran=\mo(\om)K_{i}(\om)
\ee
where we have made use of Eq.(\ref{random}). 
\be
\mo(\om) \equiv \frac{1}{-i\om+\ga[\om]}
\ee
is called the admittance and since it depends upon $\ga$ it inherits the anisotropic effect. The admittance is a 
measure of the response of the Brownian particle to external perturbations. In particular, if the external force is taken as
\be 
K_{i}(t)=K_{i}^{(0)}e^{-i\om t}
\ee
then the response is,
\be 
\lan p_{i}(t) \ran=\mo(\om)K^{(0)}_{i}e^{-i\om t}.
\ee
If the memory kernel $\ga(t-t')$ is sharply peaked around $t'=t$ then 
\be 
\int_{0}^{\infty}dt' \g^{(i)}(t-t')p_{i}(t')\approx \int_{0}^{\infty}dt' \g^{(i)}(t')p_{i}(t)=\frac{1}{t^{(i)}_{\text{relax}}}p_{i}(t).
\ee
Thus, for the generalized Langevin equation, described by Eq.(\ref{langgen}), the generalization of the relaxation time is
\be 
t^{(i)}_{\text{relax}} \sim \left( \int_{0}^{\infty}dt \ga(t)\right)^{-1}=\frac{1}{\ga[\om=0]}=\mo(\om=0).
\ee
The Wiener-Khintchine theorem relates the power spectrum $I_{\z{O}}(\om)$ of any quantity $\z{O}$ with its two-point function as follows,
\be \label{wiener}
\lan \z{O}(\om)\z{O}(\om ')\ran=2\pi \de(\om+\om ')I_{\z{O}}(\om)
\ee
where the power spectrum $I_{\z{O}}(\om)$ is defined as 
\be 
I_{\z{O}}(\om)=\int_{-\infty}^{\infty}dt\lan \z{O}(t_{0})\z{O}(t_{0}+t)\ran e^{i\om t}.
\ee
For stationary systems this does not depend upon the choice of $t_{0}$ and hence, we can as well set $t_{0}=0$. Now if we turn off 
the external force $K_{i}(t)$ then from Eq.(\ref{lanf}) we get,
\be 
p_{i}(\om)=\frac{R_{i}(\om)}{-i\om+\ga[\om]}=\mo(\om)R_{i}(\om)
\ee
which leads to the obvious result
\be \label{correlator}
I_{p_{i}}(\om)=\frac{I_{R_{i}}(\om)}{|\ga[\om]-i\om|^{2}}=|\mo(\om)|^{2}I_{R_{i}}(\om).
\ee
Making use of Eqs.(\ref{random},\ref{correlator}) we are lead to the result,
\be 
\ka^{(i)}=I_{R_{i}}=\frac{I_{p_{i}}(\om)}{|\mo(\om)|^{2}}.
\ee
The random force correlator $\ka^{(i)}$ provides yet another time scale involved in the Brownian motion. If we take $\ka^{(i)}$ to be of the form,
\be 
\ka^{(i)}(t)=\ka^{(i)}(0)e^{-\frac{t}{t_{\text{col}}}}
\ee
then $t_{\text{col}}$ is the width of the correlator. It is the temporal span over which the random forces are correlated and gives the time-scale for the duration of a collision.

In the next section, following holographic techniques prescribed in \cite{deBoer:2008gu}, we investigate the bulk realization of the boundary 
Brownian motion of a heavy probe moving in an anisotropic thermal plasma. In doing so, we first describe the profile of the 
probe string stretching between the $AdS$ boundary and the horizon as well as the black hole background dual to
the anisotropic plasma. Then we describe how to compute bulk correlators of the transverse fluctuations of the probe string.

\section{The holographic story}\label{holo}

To incorporate heavy dynamical probe quark in the boundary theory, one introduces  $N_{f}$  $D7$-flavor branes located at $r=r_{m}$. We work within probe approximation meaning $N_{f} \ll N_{c}$ and neglect the backreaction of the flavor brane on the background (for simplicity we take $N_{f}=1$). On the gauge theory side this is tantamount to working in the quenched approximation.
The probe string stretches from the boundary at $r=r_{m}$ to the black hole horizon $r=r_{h}$. The flavor brane spans the four gauge theory directions, the radial direction and also a $3$-sphere $S^{3} \subset S^{5}$. 
We take the boundary gauge theory to live at the radial coordinate $r=r_{m}$. We assume that the source of the fluctuations of the string modes
is purely Hawking radiation. Moreover, keeping the string coupling $g_{s}$ small ensures that we can  ignore the interaction between the transverse fluctuation modes and the closed string modes in the bulk.

\subsection{The anisotropic supergravity dual} \label{dual}
In this subsection we briefly provide the details of the gauge theory we are interested in and its supergravity dual. 
The gauge theory under consideration is a spatially deformed  $\z{N}=4, SU(N_{c})$ SYM plasma at 
large t'Hooft coupling $\la=g_{YM}^{2}N_{c}$. The deformation is achieved by introducing a $\t$-parameter in our theory that depends 
linearly upon any one of the three spatial directions, which we take to be $x^{3}$ in our case. Consequently, we can write the gauge 
theory action as,
\be 
S_{\text{gauge}}=S_{\text{SYM}}+\de S
\ee
where
\be 
\de S= \frac{1}{8 \pi^{2}}\int \t(x^3) \text{Tr}  F \wedge F.
\ee
The presence of $\t(=2\pi n_{D7}x^{3})$ reduces the $SO(3)$ rotational symmetry of the original theory down to a $SO(2)$ symmetry in 
the $x^{1}$-$x^{2}$ plane (where we have taken $\{t,x^{1},x^{2},x^{3}\}$ to be the gauge theory coordinates) and is responsible for making 
the theory anisotropic. Here $n_{D7}$ is a constant with energy dimension. In the context of heavy ion collisions, $x^{3}$ will 
correspond to the direction of beam whereas the $x^{1},x^{2}$-directions span the transverse plane. In heavy ion collisions the plasma will 
expand and cool down gradually and the anisotropy parameter will also decay with time. However, here we shall restrict ourselves to a 
time domain where such temporal variation can be neglected. The type IIB supergravity dual to this gauge theory was given in 
\cite{Mateos:2011ix,Mateos:2011tv} inspired by \cite{Azeyanagi:2009pr} and reads in the string frame,
\be \label{MT}
ds^{2}=r^{2}\left(-\z{FB}dt^{2}+(dx^{1})^{2}+(dx^{2})^{2}+\z{H}(dx^{3})^{2}+\frac{dr^{2}}{r^{4}\z{F}}\right)+e^{\half \phi}d\Om_{5}^{2},
\ee
\be 
\chi=ax^{3}, \qquad \phi=\phi(r)
\ee
where the axion $\chi$, is proportional to the anisotropic direction $x^{3}$, the proportionality constant $a$ being the anisotropy parameter. 
The theory also has a running dilaton $\phi(r)$. $r$ is the $AdS$ radial coordinate with the boundary at $r=\infty$ and the horizon at $r=r_{h}$, 
$d\Om_{5}^{2}$ is the metric on  the five-sphere $S^{5}$. We have suppressed  the common radius $R$ of the $AdS$ space and $S^{5}$  setting $R=1$.  
There is also a RR self-dual five-form which will not play any role in our discussion here.  The axion, which is dual to the gauge theory $\t$-term, is responsible for making the background anisotropic. It turns out \cite{Mateos:2011ix} that the anisotropy parameter $a$ is proportional to $n_{D7}$, the number density of $D7$-branes along the $x^{3}
$ direction, $a=\la n_{D7}/4 \pi N_{c}$. The $D7$-branes, which source the axion, wrap around $S^{5}$ and extend along the transverse directions, 
$x^{1},x^{2}$. However, the $D7$-branes do not span the radial direction and hence, do not reach the boundary. So they do not contribute any new 
degrees of freedom to the theory. $\z{F,B,H}$  are all functions of the radial coordinate $r$ and are known analytically only in the limiting 
cases when the anisotropy is very high or low (with respect to the temperature). In the intermediate regime, they are known only numerically. 
$\z{F}$ is the usual `blackening factor' that vanishes at the horizon, i.e., $\z{F}(r_{h})=0$. The presence of anisotropy implies that the dual 
theory develops an anisotropic horizon. The strength of anisotropy can be tuned by varying the parameter $a$. In this paper, we shall consider only 
weakly anisotropic plasma (the small $a$ or high temperature $T$ limit, whence $a/T \ll 1$). In this regime, the functions $\z{F,B,H}$ can be 
expanded to leading 
order in $a$ around the black $D3$-brane solution,
\bea \label{fn1}
\z{F}(r)&=&1-\frac{r_{h}^{4}}{r^{4}}+a^{2}\z{F}_{2}(r)+\z{O}(a^{4}),\nn \\
\z{B}(r)&=&1+a^{2}\z{B}_{2}(r)+\z{O}(a^{4}),\nn \\
\z{H}(r)&=&e^{-\phi(r)} \qquad \text{with} \qquad \phi(r)=a^{2}\phi_{2}(r)+\z{O}(a^{4})
\eea
where
\bea \label{fn2}
\z{F}_{2}(r)&=&\frac{r_{h}^{2}}{24r^{4}}\left[\frac{8\left(r^{2}-r_{h}^{2}\right)}{r_{h}^{2}}-10\log2  +\frac{3r^{4}+7r_{h}^{4}}{r_{h}^{4}}\log\left(1+\frac{r_{h}^{2}}{r^{2}}\right) \right], \nn \\
\z{B}_{2}(r)&=&-\frac{1}{24 r _{h}^{2}}\left[\frac{10r_{h}^{2}}{r_{h}^{2}+r^{2}}+\log\left(1+\frac{r_{h}^{2}}{r^{2}}\right)\right], \nn \\
\phi_{2}(r)&=&-\frac{1}{4r_{h}^{2}}\log\left(1+\frac{r_{h}^{2}}{r^{2}}\right).
\eea
The Hawking temperature of the above solution is
\be 
T=\frac{r_{h}}{\pi }+\frac{a^{2}}{r_{h}}\frac{(5\log2-2)}{48\pi}+\z{O}(a^{4})
\ee
which is  identified as the temperature of the deformed SYM theory. The horizon position can be obtained in terms of the temperature, 
which, in the limit $a/T \ll 1$, reads
\be  \label{temp}
r_{h} \sim \pi T\left[ 1-a^{2}\frac{5\log2-2}{48\pi^{2}T^{2}}\right]+\z{O}(a^{4}).
\ee

\subsection{Bulk view of  Brownian motion}\label{bulk}
To study the dynamics of the fundamental string in the background given by Eq.(\ref{MT}) we need to evaluate the Nambu-Goto string worldsheet 
action,
\be \label{NG}
S_{NG}=\frac{1}{2\pi \a'}\int d\si d\tau \sqrt{-\text{det}g_{\a \b}}
\ee
where $g_{\a \b}$ is the induced metric on the string worldsheet,
\be 
g_{\a \b}=G_{\mu \nu}\frac{\pa X^{\mu}}{\pa \xi^{\a}}\frac{\pa X^{\nu}}{\pa \xi^{\b}}.
\ee
Here $\xi^{\a,\b}$ are the coordinates on the string worldsheet $\Si$: $\xi^{0}=\tau$ and $\xi^{1}=\si$, $G_{\mu \nu}$ is the 
ten-dimensional metric as given in Eq.(\ref{MT}) and $\{X^{\mu}(\tau,\si)\}$ are the ten-dimensional coordinates  which specify the 
string embedding in the full ten-dimensional spacetime. We choose the static gauge for evaluating Eq.(\ref{NG}) as $\tau=t,\si=r$. 
The trivial solution that satisfies the equation of motion obtained by variation of $S_{NG}$ is given by $X^{m}=\{t,\vec{0},r \}$. 
This corresponds to a quark that is in equilibrium in a thermal bath and in the bulk picture to a string hanging straight down radially. 
We now wish to consider fluctuations around this classical solution. We want to see the effects of anisotropy both along the anisotropic 
direction as well as in the isotropic plane. To this end we consider fluctuations of the form: $X^{m}=\{t,X_{1}(t,r),0,X_{3}(t,r),r \}$ 
where $X_{1}(t,r)$ is a fluctuation in a isotropic direction while $X_{3}(t,r)$ is a perturbation 
along the anisotropic direction. The position of the quark is given by, $x^{\mu}=\{t,X_{1}(t,r_{m}),0,X_{3}(t,r_{m}) \}$. 
Using this parametrization we find out the components of the worldsheet metric as,
\be
\begin{aligned}
 g_{\tau \tau}&=r^{2}\left(-\z{FB}+(\dot{X}_{1})^{2}+\z{H}(\dot{X}_{3})^{2} \right), \\
 g_{\si \si}&=r^{2}\left((X_{1}')^{2}+\z{H}(X_{3}')^{2}+\frac{1}{r^{4}\z{F}} \right),\\
 g_{\tau \si}&=r^{2}\left(\z{H}\dot{X_{1}}X_{3}'+\z{H}X_{1}'\dot{X_{3}}\right)
\end{aligned}
\ee
where $X_{i}' \equiv \partial_{\si}X_{i}$ and $\dot{X_{i}} \equiv \partial_{\tau}X^{i}$. From now on, we suppress the explicit 
$r$-dependence of the metric elements $\z{F,B,H}$. If we restrict ourselves to small perturbation around the classical solution we 
can safely leave out terms higher than quadratic order in the fluctuations whence the action reduces to \footnote{This essentially 
means that we are in the regime $|\partial_{t}X_{i}|\ll 1$ which, in turn, implies taking the nonrelativistic limit. 
Hence, on the gauge theory side, the dual picture will also be nonrelativistic.}
\be \label{NGm}
S_{NG}=\frac{1}{4\pi \a'}\int d\tau d\si \sqrt{\z{B}}\left[\z{F}r^{4}\left((X_{1}')^{2}+\z{H}(X_{3}')^{2}\right)-\frac{1}{\z{FB}}\left((\dot{X_{1}})^{2}+\z{H}(\dot{X_{3}})^{2} \right) \right].
\ee
While writing Eq.(\ref{NGm}) we have omitted a constant factor that is independent of $X_{i}$. Variation of the above action
 yields the equation of motion for the fluctuation $X_{3}$
\begin{subequations}
\be \label{eom3}
\ddot{X_{3}}-\frac{\z{F}\sqrt{\z{B}}}{\z{H}}r_{h}^{2}\partial_{y}\left( \sqrt{\z{B}}\z{HF}y^{4}X_{3}'\right)=0
\ee
where we have used the new scaled coordinate, $y=r/r_{h}$ and now the prime $'$ denotes derivative with respect to $y$. 
The equation of motion for $X_{1}$ is obtained in a similar fashion,
\be \label{eom1}
\ddot{X_{1}}-\z{F}\sqrt{\z{B}}r_{h}^{2}\partial_{y}\left( \sqrt{\z{B}}\z{F}y^{4}X_{1}'\right)=0
\ee
\end{subequations}
which is the same as Eq.(\ref{eom3}) with $\z{H}=1$. Later on, we shall also consider forced motion of the quark under the effect of 
an electromagnetic field. This is simply achieved by switching on a $U(1)$ electromagnetic field on the flavor $D7$-brane. Since the string 
end point on the boundary represents a quark, it is charged, and hence will couple to the electromagnetic field. Consequently, 
we need to incorporate this effect at the level of the action. The action $S_{NG}$ is then generalized to $S=S_{NG}+S_{b}$ where
\be 
S_{b}=\int_{\partial \Si}\left(A_{t}+A_{i}\dot{X_{i}} \right)dt.
\ee
Since it is just a boundary term it will not affect the dynamics of the string in the bulk. However, it will modify the boundary 
conditions that we need to impose upon the string endpoint. We need to find solutions to Eqs.(\ref{eom3},\ref{eom1}) 
near the boundary which we shall do by employing the matching technique. The solutions are, in general, quite complicated. 
However, they are readily obtained near the horizon. So before finding out the actual solutions let us see how these solutions behave 
in the vicinity of $y \rightarrow 1$. First of all, we inflict a coordinate transformation $r \rightarrow r_{*}$  which takes us 
to the tortoise coordinates so that
\be 
\frac{d}{dr}=\frac{1}{r^{2}\z{F}\sqrt{\z{B}}}\frac{d}{dr_{*}}
\ee
and 
\be \label{tortoise}
dr=r^{2}\z{F}\sqrt{\z{B}}dr_{*}.
\ee
In this new coordinate system, the Nambu-Goto action assumes the form,
\be 
S_{NG}=\frac{1}{4\pi \a'}\int d\tau dr_{*}r^{2}\left[\left((\partial_{r_{*}}X_{1})^{2}-(\dot{X_{1}})^{2}\right) +\z{H}\left(    (\partial_{r_{*}}X_{3})^{2}-(\dot{X_{3}})^{2}\right)\right].
\ee
Near the horizon it simplifies to,
\be 
S_{NG}=\frac{1}{4 \pi \a'}r_{h}^{2}\int d\tau dr_{*}\left[\left((\partial_{r_{*}}X_{1})^{2}-(\dot{X_{1}})^{2}\right) +\z{H}(r_{h})\left(    (\partial_{r_{*}}X_{3})^{2}-(\dot{X_{3}})^{2}\right) \right].
\ee
The equation of motion for both $X_{1}$ and $X_{3}$ obtained by varying this action turns out to be the same,
\be \label{nearhorizoneq}
\left( \partial^{2}_{r_{*}}-\partial^{2}_{\tau}\right)X_{1,3}=0.
\ee
So near the boundary, the fluctuations are governed by a Klein-Gordon equation for massless scalars. From now on, in this section, 
we shall  refer to the fluctuations as $X_{i}$, it being understood that everything we discuss here holds true for both $X_{1}$ 
as well as $X_{3}$. From Eq.(\ref{MT}) it is clear that $t$ is an isometry of the background and hence we can try solutions of the form,
\be \label{ans}
X_{i}(t,r)\sim e^{-i\om t}g_{\om}(r)
\ee
Eq.(\ref{nearhorizoneq}) has two independent  solutions corresponding to ingoing and outgoing waves respectively which we write as,
\begin{subequations}
 \be 
 X_{i}^{\text{out}}(r)=e^{-i\om t}g_{i}^{\text{out}}(r)\sim e^{-i\om(t-r_{*})}
 \ee
 \be 
 X_{i}^{\text{in}}(r)=e^{-i\om t}g_{i}^{\text{in}}(r)\sim e^{-i\om(t+r_{*})}.
 \ee
\end{subequations}
To find $r_{*}$ we need to solve Eq.(\ref{tortoise}) which yields,
\be 
r_{*}=\frac{1}{4r_{h}}\log\left(\frac{r}{r_{h}}-1\right)\left[1-\frac{\tilde{a}^{2}}{48}(5\log2-2) \right]
\ee
where we have defined $\tilde{a}=\frac{a}{r_{h}}\sim \frac{a}{\pi T}$.
Hence, 
\be \label{nearhorizonsol}
g_{i}^{\text{out/in}}(r)=\left(\frac{r}{r_{h}}-1\right)^{\pm \frac{i \nu}{4}\left(1-\frac{\tilde{a}^{2}}{48}(5\log2-2) \right)}
\ee
where $\nu=\frac{\om}{r_{h}}$. One thus finds that, $g_{i}^{\text{out}}=(g_{i}^{\text{in}})^{*}$.\\
Following standard quantization techniques of scalar fields in curved spacetime we can perform a mode expansion of the fluctuations as
\be 
X_{i}(t,r)=\int_{0}^{\infty}\frac{d\om}{2\pi}[a_{\om}u_{\om}(t,r)+a^{\dagger}_{\om}u_{\om}(t,r)^{*}].
\ee
Here $u_{\om}(t,r)$ is a set of positive frequency basis. These modes can in turn be expressed as a linear combination of the ingoing and 
the outgoing waves
\be 
u_{\om}(t,r)=A[g^{\text{out}}(r)+B g^{\text{in}}(r)]e^{-i\om t}.
\ee
The constant $B$ is determined by imposing boundary condition at $r=r_{m}$, i.e., $y=1$. However, as we shall later see, $B$ turns to be 
a pure phase. This implies that the outgoing and the ingoing modes have the same amplitude. This signifies that the black hole environment 
which can emit Hawking radiation is in a state of thermal equilibrium. One is then left with determining the constant $A$ which is fixed by 
demanding normalization of the modes through the conventional Klein-Gordon inner product defined via,
\be 
(f_{i},g_{j})_{\si}=-\frac{i}{2\pi \a'}\int_{\si}\sqrt{\tilde{g}}n^{\mu}G_{ij}(f_{i}\partial_{\mu}g_{j}^{*}-\partial_{\mu}f_{i}g_{j}^{*}).
\ee
Here, $\si$ defines a Cauchy surface in the $(t,r)$ subspace of the ten-dimensional spacetime metric, $\tilde{g}$ is the induced metric on the 
surface $\si$ and $n^{\mu}$ denotes a unit normal to $\si$ in the future direction. Without any loss of generality we can take the surface 
$\si$ to be a constant $t$ surface since the inner product does not depend upon the exact choice of the surface in the ($t,r$)-plane \cite{BD}. 
Following \cite{Atmaja:2010uu} we argue that the primary contribution to the above integral arises from the IR region. Of course, regions away 
from the horizon do contribute but since the horizon is semi-infinite in the tortoise coordinate, the normalization is completely fixed by the 
near-horizon regime. For the anisotropic direction this gives,
\be 
(f_{i},g_{j})_{\si}=-\frac{i\de_{ij}r_{h}^{2}\z{H}(r_{h})}{2 \pi \a'}\int \limits_{r_{*}\rightarrow -\infty}dr_{*}(f_{i}\dot{g_{j}}^{*}-\dot{f_{i}}g_{j}^{*})
\ee
from which we can extract $A$ to be,
\be \label{normalise}
A=\sqrt{\frac{\pi \a'}{\om r_{h}^{2}\z{H}(r_{h})}}.
\ee
On the other hand, for fluctuations along the isotropic direction we have,
\be 
(f_{i},g_{j})_{\si}=-\frac{i\de_{ij}r_{h}^{2}}{2 \pi \a'}\int\limits_{r_{*}\rightarrow -\infty}dr_{*}(f_{i}\dot{g_{j}}^{*}-\dot{f_{i}}g_{j}^{*})
\ee
which fixes $A$ as,
\be \label{normaliseiso}
A=\sqrt{\frac{\pi \a'}{\om r_{h}^{2}}}.
\ee
The  normalisation ensures that the inner product $(u_{\om},u_{\om})=1$ which in turn, guarantees that the canonical commutation relations 
are satisfied,
\be 
[a_{\om},a_{\om'}]=[a^{\dagger}_{\om},a^{\dagger}_{\om'}]=0, \qquad [a_{\om},a^{\dagger}_{\om'}]=2\pi \de(\om+\om').
\ee
In the semiclassical approximation the string modes are thermally excited by the Hawking radiation of the worldsheet horizon and obey the 
Bose-Einstein distribution,
\be 
\lan a_{\om}a^{\dagger}_{\om} \ran=\frac{2\pi \de(\om+\om')}{e^{\b \om}-1}.
\ee
Equipped with this much machinery we are now ready to compute the displacement squared for the test quark in the boundary. 
This is required if we wish to find out an expression for the diffusion constant. Recalling that  the position of the Brownian particle 
is specified by $x_{i}(t)=X_{i}(t,r_{m})$ we have
\be 
\lan x_{i}(t)x_{i}(0)\ran=\int_{0}^{\infty}\frac{d\om d\om'}{(2\pi)^{2}}[\lan a_{\om}a^{\dagger}_{\om'}\ran u_{\om}(t,r_{m})u_{\om'}(0,r_{m})^{*}+\lan a_{\om}^{\dagger}a_{\om'}\ran u_{\om}(t,r_{m})^{*}u_{\om'}(0,r_{m})]
\ee
However, this is afflicted by a divergence that can be attributed to the zero point energy which persists even we go to the zero temperature limit. The way to bypass this catastrophe is to invoke the normal ordering of products
\be 
\lan :x_{i}(t)x_{i}(0):\ran=\int_{0}^{\infty}\frac{d\om}{2\pi}\frac{2|A|^{2}\cos \om t}{e^{\b \om}-1}|g^{\text{out}}(r_{m})+Bg^{\text{in}}(r_{m})|^{2}.
\ee
Finally, after a little algebra we arrive at the expression for displacement squared,
\be \label{disp}
s_{i}^{2}(t)\equiv \lan :[x_{i}(t)-x_{i}(0)]^{2}:\ran=\frac{4}{\pi}\int_{0}^{\infty} d\om |A|^{2}\frac{\sin^{2}\om t/2}{e^{\b \om}-1}|g^{\text{out}}(r_{m})+Bg^{\text{in}}(r_{m})|^{2}.
\ee
With the general formalism in place, we are now in a position to take up the problem of analyzing Brownian motion in an anisotropic 
strongly coupled plasma from the holographic point of view. In \ref{ani} we study the case of Brownian motion in the plasma along the 
anisotropic direction. We discuss this case in detail. Later in \ref{iso} we consider Brownian motion along one of the isotropic directions.

\subsection{Brownian motion along anisotropic direction}\label{ani}
Our first job will be to solve Eq.(\ref{eom3}) in the asymptotic limit. Making use of Eq.(\ref{ans}) we recast Eq.(\ref{eom3}) as,
\be 
\nu^{2}g(y)+\frac{\z{F}\sqrt{\z{B}}}{\z{H}}\partial_{y}\left( \sqrt{\z{B}}\z{HF}y^{4}g'\right)=0.
\ee
Inserting the explicit expressions of the various functions, this can be written as,
\be \label{mastereq}
g''(y)+4\frac{y^{3}}{y^{4}-1}\left[1+\tilde{a}^{2}\Psi(y) \right]g'(y)+\frac{y^{4}\nu^{2}}{(y^{4}-1)^{2}}\left[1+\tilde{a}^{2}\Upsilon(y) \right]g(y)=0
\ee
where
\be \label{mastercoeff}
\begin{aligned} 
& \Psi(y)= \frac{1}{96y^{4}(y^{4}-1)}\left[3-9y^{2}-23y^{6}+y^{4}(29+40\log2)-40y^{4}\log\left(1+\frac{1}{y^{2}}\right) \right] \\ 
& \Upsilon(y)= \frac{1}{24(y^{4}-1)}\left[ 6-6y^{2}+20\log2-5(3+y^{4})\log\left(1+\frac{1}{y^{2}}\right)\right].
\end{aligned}
\ee
We need to find a solution to this equation. However, as it turns out, obtaining an analytic solution is a notoriously difficult 
problem for any arbitrary frequency $\nu$. To circumvent this difficulty we work only in the low frequency approximation and then 
attempt to solve the equation by the `matching technique'. Since we only require the solution near the boundary, we just give here 
the expression of the required solution. The interested reader is referred to appendix \ref{sol} for the details of the solution. 
We shall have two solutions corresponding to the ingoing and outgoing waves
\be \label{asymsol} 
g^{\text{out/in}}=k_{1}^{\text{out/in}}\left[1+\frac{\nu^{2}}{2y^{2}}+\z{O}\left(\frac{1}{y^{4}}\right) \right]+k_{3}^{\text{out/in}}\left[\frac{1}{y^{3}}+\z{O}\left(\frac{1}{y^{5}}\right)\right]
\ee
where
\be
\begin{aligned}
k_{1}^{\text{out/in}}= &1\mp \frac{i\nu}{8}\left(\pi -2\log2 \right) \pm \frac{i\nu\tilde{a}^{2}}{768}\left[28-16 \b(2) -20(\log2)^{2}\right.\\
& \left.+\pi(-8+\pi+14\log2)+8\log2  \right]+ \z{O}(\nu^{2})\\
k_{3}^{\text{out/in}}=& \mp \frac{i\nu}{3}(1+\frac{\tilde{a}^{2}}{4}\log2)+\z{O}(\nu^{2})
\end{aligned}
\ee
where $\b(2)\footnote{$\b(s)$ is the Dirichlet beta function given by $\b(s)=\sum \limits_{n=0}^{\infty}\frac{(-1)^n}{(2n+1)^{s}}$.}\sim 0.915 966$.
We find that the relation, $g^{\text{out}}=g^{\text{in}*}$, obtained earlier in the near horizon analysis, continues to hold true in the 
asymptotic limit.

We can now use these solutions, supplemented by the appropriate boundary conditions to find out various quantities of interest. However, 
before going into the intricacies of the actual computation, let us digress a little bit to clarify the boundary conditions involved 
in the problem.

Although we are interested in the worldsheet theory of the probe string, the choice of the static gauge implies that the characteristics of the background spacetime is encoded in the induced metric. Hence, we can exploit the rules of the $AdS/CFT$ correspondence to understand the boundary conditions. When working in the Lorentzian $AdS/CFT$ it is customary to choose normalisable boundary conditions \cite{Balasubramanian:1998sn} for the modes. In the present scenario this amounts to pushing the boundary all the way up to $y \rightarrow \infty$. However, the $AdS/CFT$ dictionary  tells us that the radial distance  is mapped holographically to the mass of the probe quark so that placing the boundary at $y \rightarrow \infty$ essentially means that we are considering our probe quark to be infinitely massive. Of course, this at once rules out any possibility of the quark undergoing Brownian motion. The problem can be solved if, instead, we impose a UV cut-off in our theory. More specifically, we introduce a UV 
cut-off 
surface and identify it with the boundary where the gauge theory lives. In fact, this is exactly the location of the flavor brane $y_{m}$ to which the endpoint of the string is attached. The relation between the position of the UV cut-off and the mass of the probe can be read off easily as,
\be 
m=\frac{1}{2\pi \a'}\int _{r_{h}}^{r_{m}}dr\sqrt{-g_{tt}g_{rr}}=\frac{1}{2\pi \a'}\left[y_{m}-1+\frac{\tilde{a}^{2}}{24}\left(\log2-3\pi \right)\right]
\ee
and the worldsheet metric elements $g_{tt},g_{rr}$ are written for the classical string configuration, i.e., omitting the contribution arising out of the fluctuations. On this surface we can impose Neumann boundary condition\footnote{One can not impose Dirichlet condition since it implies no fluctuation on the boundary at all.} $\partial_{r}X_{i}=0$. However, this works only when we consider the free Brownian motion of the particle in the absence of any external force.  In the case of forced motion this is modified to,
\be \label{genbc}
\Pi^{y}_{i}\big|_{\partial \Si}\equiv \frac{\partial \z{L}}{\partial X_{i}'}=K_{i}=K_{i}^{(0)}e^{-i\om t}.
\ee
where we have assumed a fluctuating external force.

Now the general solution $X_{i}$ is a linear combination of the outgoing and the ingoing modes at the horizon,
\be  \label{gensol}
X_{i}=A^{\text{out}}X_{i}^{\text{out}}+A^{\text{in}}X_{i}^{\text{in}}.
\ee
where $X_{i}^{\text{out/in}}=e^{-i \om t}g^{\text{out/in}}$ and $g^{\text{out/in}}$ is given in Eq.(\ref{asymsol}). In the semiclassical 
approximation the outgoing modes are thermally excited by the Hawking radiation emanating from the black hole whereas the ingoing modes can 
be arbitrary. Since the Hawking radiation is a random phenomena the phase of $A^{\text{out}}$ takes random values and and its average 
$\lan A^{\text{out}}\ran$ vanishes. So we can omit the first term in Eq.(\ref{gensol}) and need to consider only the ingoing wave.\\
When one plugs in the form of the Lagrangian in Eq.(\ref{genbc}) one finds that, like the equations of motion, the boundary conditions 
along the anisotropic direction and the isotropic directions decouple which allows us to treat each direction separately. 
Coming back to the particular case of the anisotropic direction, the boundary condition given in Eq.(\ref{genbc}) assumes the form,
\be 
\frac{1}{2 \pi \a'}\z{HF}\sqrt{\z{B}}y^{4}r_{h}^{3}X_{3}'\big{|}_{y=y_{m}}=K_{3}=K^{(0)}_{3}e^{-i\om t}.
\ee
This yields
\be 
A^{\text{in}}=\left.\frac{2\pi \a' K_{3}^{(0)}}{\z{HF}\sqrt{\z{B}}y^{4}r_{h}^{3}g'(y)}\right|_{y=y_{m}}.
\ee
where $g(y)$ represents the ingoing solution in Eq.(\ref{asymsol}). So, on the boundary the average position of the Brownian quark is given by,
\be 
\lan x_{3}(t)\ran=\left.\lan X_{3}(t,y_{m})\ran=K^{(0)}_{3}e^{-i\om t}\frac{2\pi \a'g}{\z{HF}\sqrt{\z{B}}y^{4}r_{h}^{3}g'}\right|_{y=y_{m}}.
\ee
The average momentum is,
\be 
\lan p_{3}(t)\ran=\left.m\lan \dot{x_{3}}\ran=-K_{3}\frac{2i\pi \a' m \nu g}{\z{HF}\sqrt{\z{B}}y^{4}r_{h}^{2}g'}\right|_{y=y_{m}}.
\ee
Comparison with Eq.(\ref{adm}) results in,
\be 
\mu^{||}(\nu)\equiv \mu^{(3)}(\nu)=\left.-\frac{2i\pi \a' m \nu g}{\z{HF}\sqrt{\z{B}}y^{4}r_{h}^{2}g'}\right|_{y=y_{m}}.
\ee
Here we have used the superscript ``$||$'' to denote quantities along the anisotropic direction (the $x_{3}$ direction). 
Reinstating the expressions for the various functions and expanding up to $\z{O}(\tilde{a}^{2})$ in the low frequency regime we obtain the relaxation time for heavy quark diffusing along the anisotropic direction,
\be \label{parmu}
\mu^{||}(0)=t^{||}_{\text{relax}}=\frac{2m}{\pi \sqrt{\la}T^{2}}\left[ 1-\frac{a^{2}}{24 \pi^{2}T^{2}}(2+\log2) \right]
\ee
from which one gets the drag coefficient along the anisotropic direction
\be \label{pardrag}
\g^{||}[0]=\frac{\pi \sqrt{\la}T^{2}}{2m}\left[1+\frac{a^{2}}{24 \pi^{2}T^{2}}(2+\log2) \right]=\g_{iso}\left[1+\frac{a^{2}}{24 \pi^{2}T^{2}}(2+\log2) \right]
\ee
where $\g_{iso}$ represents the drag coefficient when the quark moves in a isotropic  SYM plasma. Here we have used the standard $AdS/CFT$ dictionary, $R^{4}=(\a')^{2}\la$ with $R=1$ in our convention. Our expression for the friction coefficient $\g^{||}$ matches exactly with that obtained in \cite{Chernicoff:2012iq} in the nonrelativistic limit $v\ll1$ along the anisotropic direction. Note that the drag force increases compared to its isotropic counterpart when the quark moves along the anisotropic direction.\\
Next we turn towards computing the displacement squared for the Brownian particle from which we can extract the expression for the diffusion 
constant $D^{||}$. We have already provided a generic expression for $s_{i}^{2}$ in Eq.(\ref{disp}). The details of the calculation will 
depend upon the background metric. Let us again return to the boundary condition Eq.(\ref{genbc}), but now with the gauge fields 
turned off. Eq.(\ref{genbc}) then reads for the anisotropic direction,
\be 
\left. \frac{\partial \z{L}}{\partial X_{3}'}=\frac{1}{2 \pi \a'}\z{HF}\sqrt{\z{B}}y^{4}r_{h}^{3}X_{3}'\right|_{y=y_{m}}=0
\ee
which translates to $X_{3}'=0$ at the boundary. The fluctuations $X_{i}(t,y)$ can be expressed as the sum of outgoing and ingoing modes as,
\be \label{linearcomb}
X_{i}(t,y)=A[g^{\text{out}}(y)+B g^{\text{in}}(y)]e^{-i\om t}.
\ee
It then easily follows that, $X_{3}'=0$ implies
\be 
B=\left.-\frac{g^{\text{out}'}}{g^{\text{in}'}}\right|_{y=y_{m}}=1+\z{O}(\nu)
\ee
which gives,
\be \label{sum}
|g^{\text{out}}(y_{m})+Bg^{\text{in}}(y_{m})|^{2}=4+\z{O}(\nu).
\ee
Using Eqs.(\ref{normalise},\ref{sum}) in Eq.(\ref{disp}) one then has,
\be 
s_{3}^{2}= \frac{4t}{\pi T \sqrt{\la}}\left[1-\frac{a^{2}}{24 \pi^{2}T^{2}}(2+\log2) \right].
\ee
Hence, the diffusion constant along the anisotropic direction is,
\be 
D^{||}= \frac{2}{\pi T \sqrt{\la}}\left[1-\frac{a^{2}}{24 \pi^{2}T^{2}}(2+\log2) \right]=\frac{T}{m\g^{||}}.
\ee
This is nothing but the Einstein-Sutherland relation (Eq.(\ref{einsuth})) mentioned earlier. We have thus performed an explicit verification 
of the relation from the bulk point of view. \\
Finally, we proceed to verify the fluctuation-dissipation theorem for which we need to know the random force correlator. First of all, 
we compute the two-point correlator of the momentum along the $i$-th direction,
\be
\begin{aligned}
 \lan:p_{i}(t)p_{i}(0):\ran & \equiv -m^{2}\partial_{t}^{2}\lan:x_{i}(t)x_{i}(0):\ran\\
 &=\int_{0}^{\infty}\frac{d\om}{2\pi}\frac{2m^{2}\om^{2}|A|^{2}\cos \om t}{e^{\b \om}-1}|g^{\text{out}}(y_{m})+Bg^{\text{in}}(y_{m})|^{2}.
\end{aligned}
\ee
Invoking the Wiener-Khintchine theorem (Eq.(\ref{wiener})), and the expression for $A$ (Eq.(\ref{normalise}))and specialising to 
the anisotropic direction we find,
\be 
I_{p_{3}}(\om)=4\frac{m^{2}\pi}{r_{h}^{2}\a'\z{H}(y=1)\b}\frac{\b \om}{e^{\b \om}-1}.
\ee
Expanding in $\om$ and keeping only the leading order term one has
\be 
I_{p_{3}}(\om)=\frac{4m^{2}}{\sqrt{\la}\pi T}\left( 1+\frac{a^{2}}{24\pi^{2}T^{2}}(5\log2-2)\right)\left(1-\frac{a^{2}}{4\pi^{2}T^{2}}\log2 \right)+\z{O}(\om).
\ee
Now, the Langevin coefficient along the direction of anisotropy is
\be
\begin{aligned}
\ka^{||}&=I_{R_{3}}=\frac{I_{p_{3}}(\om)}{|\mu^{||}(\om)|^{2}}\\
&=2mT\frac{\pi \sqrt{\la}T^{2}}{2m}\left[1+\frac{a^{2}}{24\pi^{2}T^{2}}(2+\log2) \right]=2mT\g^{||}\\
&=\ka_{iso}\left[1+\frac{a^{2}}{24\pi^{2}T^{2}}(2+\log2) \right]
\end{aligned}
\ee
(where $\ka_{iso}$ is the Langevin coefficient in isotropic plasma) which is nothing but the statement of the fluctuation-dissipation theorem. We thus observe that the strength of the auto-correlator along the anisotropic direction increases in the presence of anisotropy. Thus, we explicitly check the validity of the fluctuation-dissipation theorem for a heavy test quark executing Brownian motion in a strongly coupled, anisotropic plasma when the fluctuations are aligned with the direction of anisotropy.

\subsection{Brownian motion transverse to the anisotropic direction}\label{iso}
In this subsection we discuss the case of the Brownian motion in the isotropic plane. For definiteness, we take the motion to 
be along $X_{1}$ direction. The calculations in this case proceeds in almost the same way as in $\ref{ani}$. As is evident 
upon comparing Eqs.(\ref{eom3},\ref{eom1}) the equation of motion in the isotropic direction can be simply 
obtained by setting $\z{H}=1$ in the anisotropic case. This can also be understood by looking at the metric in Eq.(\ref{MT}). 
So we shall be brief in our discussion here. The equation to solve is
\be 
\nu^{2}g(y)+\z{F}\sqrt{\z{B}}\partial_{y}\left( \sqrt{\z{B}}\z{F}y^{4}g'\right)=0
\ee
which can be recast as,
\be
g''(y)+4\frac{y^{3}}{y^{4}-1}\left[1+\tilde{a}^{2}\tilde{\Psi}(y) \right]g'(y)+\frac{y^{4}\nu^{2}}{(y^{4}-1)^{2}}\left[1+\tilde{a}^{2}\tilde{\Upsilon}(y) \right]g(y)=0
\ee
where
\be \label{mastercoeffiso}
\begin{aligned} 
& \tilde{\Psi}(y)= \frac{1}{96y^{4}(y^{4}-1)}\left[15-21y^{2}-11y^{6}+y^{4}(17+40\log2)-40y^{4}\log\left(1+\frac{1}{y^{2}}\right) \right] \\ 
& \tilde{\Upsilon}(y)= \frac{1}{24(y^{4}-1)}\left[ 6-6y^{2}+20\log2-5(3+y^{4})\log\left(1+\frac{1}{y^{2}}\right)\right].
\end{aligned}
\ee
As in the anisotropic version, here, too, we look for solutions by resorting to the matching technique. Here we present only the final form of the solution in the asymptotic limit,
\be \label{asymsoliso} 
g^{\text{out/in}}=\tilde{k}_{1}^{\text{out/in}}\left[1+\frac{\nu^{2}}{2y^{2}}+\z{O}\left(\frac{1}{y^{4}}\right) \right]+\tilde{k}_{3}^{\text{out/in}}\left[\frac{1}{y^{3}}+\z{O}\left(\frac{1}{y^{5}}\right)\right]
\ee
where
\be
\begin{aligned}
\tilde{k}_{1}^{\text{out/in}}= &1\mp \frac{i\nu}{8}\left(\pi -2\log2 \right) \mp \frac{i\nu\tilde{a}^{2}}{768}\left[-80 \b(2)\right.\\
& \left. +\pi(8+5\pi)-4(7+2\log2)+10(\pi+2\log2)\log2  \right]+ \z{O}(\nu^{2})\\
\tilde{k}_{3}^{\text{out/in}}=& \mp \frac{i\nu}{3}.
\end{aligned}
\ee
We thus find that while the $y$-dependence is the same as in its anisotropic counterpart; only the coefficients $\tilde{k}_{1}$ 
 and $\tilde{k}_{3}$ are different. Note that in particular, the coefficient $\tilde{k}_{3}$ does not pick up any contribution 
from anisotropy. The boundary condition now reads in the presence of the gauge field on the boundary
\be 
\frac{1}{2 \pi \a'}\z{F}\sqrt{\z{B}}y^{4}r_{h}^{3}X_{1}'\big{|}_{y=y_{m}}=K_{1}=K^{(0)}_{1}e^{-i\om t}.
\ee
which fixes the normalisation factor
\be 
A^{\text{in}}=\left.\frac{2\pi \a' K_{1}^{(0)}}{\z{F}\sqrt{\z{B}}y^{4}r_{h}^{3}g'}\right|_{y=y_{m}}.
\ee
One can now easily obtain expressions for the position and hence, the momentum of the Brownian quark from which follows the expression for the admittance,
\be 
\mu^{\perp}(\nu)=\left.-\frac{2i\pi \a' \nu m g}{\z{F}\sqrt{\z{B}}y^{4}r_{h}^{2}g'}\right|_{y=y_{m}}.
\ee
with $g(y)$ now being the ingoing solution in Eq.(\ref{asymsoliso}). Here we denote the direction transverse to the anisotropic one as ``$\perp$''. Reinstating the expressions for the various functions and expanding upto $\z{O}(\tilde{a}^{2})$ in the low frequency domain we obtain the relaxation time for fluctuations in the transverse plane.
\be \label{perpmu}
\mu^{\perp}(0)=t^{\perp}_{\text{relax}}=\frac{2m}{\pi \sqrt{\la}T^{2}}\left[ 1+\frac{a^{2}}{24 \pi^{2}T^{2}}(5\log2-2) \right]
\ee
from which one gets the drag coefficient along the isotropic direction
\be \label{perpdrag}
\g^{\perp}[0]=\frac{\pi \sqrt{\la}T^{2}}{2m}\left[1-\frac{a^{2}}{24 \pi^{2}T^{2}}(5\log2-2) \right]=\g_{iso}\left[1-\frac{a^{2}}{24 \pi^{2}T^{2}}(5\log2-2) \right].
\ee
This expression for the friction coefficient $\g^{\perp}$ in the isotropic direction agrees with that obtained in \cite{Chernicoff:2012iq} in the nonrelativistic limit $v\ll1$. It is to be observed, that the isotropic direction also picks up  correction from anisotropy, i.e., even the isotropic plane can ``feel'' the presence of anisotropy in the normal direction. Moreover, while the presence of anisotropy increases the drag force along the anisotropic direction it leads to a suppression in the drag force in the isotropic plane.\\
The computation for the displacement squared for the Brownian particle proceeds in exactly similar fashion as in the previous subsection. 
Switching off the external field we impose the free Neumann condition,
\be 
\left. \frac{\partial \z{L}}{\partial X_{i}'}=\frac{1}{2 \pi \a'}\z{F}\sqrt{\z{B}}y^{4}r_{h}^{3}X_{1}'\right|_{y=y_{m}}=0
\ee
which translates to $X_{1}'=0$ at the boundary that furnishes,
\be 
B=\left.-\frac{g^{\text{out}'}}{g^{\text{in}'}}\right|_{y=y_{m}}=1+\z{O}(\nu)
\ee
which implies,
\be \label{sumiso}
|g^{\text{out}}(y_{m})+Bg^{\text{in}}(y_{m})|^{2}=4+\z{O}(\nu).
\ee
Using Eqs.(\ref{sumiso},\ref{normaliseiso}) in Eq.(\ref{disp}) one then has,
\be 
s_{1}^{2}= \frac{4t}{\pi T \sqrt{\la}}\left[1+\frac{a^{2}}{24 \pi^{2}T^{2}}(5\log2-2) \right].
\ee
We can now easily read off the diffusion constant to be,
\be \label{perpdiff}
D^{\perp}= \frac{2}{\pi T \sqrt{\la}}\left[1+\frac{a^{2}}{24 \pi^{2}T^{2}}(5\log2-2) \right].
\ee
A comparison of Eqs.(\ref{perpdrag},\ref{perpdiff}) reveals the relation,
\be 
D^{\perp}=\frac{T}{m\g^{\perp}}
\ee
which verifies the validity of the Einstein-Sutherland relation in the isotropic plane.\\
Next we find the random force correlator  $\ka^{\perp}$ along the isotropic direction.
We have,
\be
\begin{aligned}
I_{p_{1}}(\om)&=4\frac{m^{2}\pi}{r_{h}^{2}\a'\b}\frac{\b \om}{e^{\b \om}-1}\\
&=\frac{4m^{2}}{\sqrt{\la}\pi T}\left( 1+\frac{a^{2}}{24\pi^{2}T^{2}}(5\log2-2)\right)+\z{O}(\om).
\end{aligned}
\ee
Now,
\be
\begin{aligned}
\ka^{\perp}&=I_{R_{1}}=\frac{I_{p_{1}}(\om)}{|\mu^{\perp}(\om)|^{2}}\\
&=2mT\frac{\pi \sqrt{\la}T^{2}}{2m}\left[1-\frac{a^{2}}{24\pi^{2}T^{2}}(5\log2-2) \right]=2mT\g^{\perp}\\
&=\ka^{iso}\left[1-\frac{a^{2}}{24\pi^{2}T^{2}}(5\log2-2) \right].
\end{aligned}
\ee
Hence, we find that the fluctuation-dissipation theorem continues to hold true in the isotropic plane too and also the random forces are less correlated in the isotropic plane due to the presence of anisotropy in the perpendicular direction.

\section{Conclusion}\label{con}
In this work we have studied the holographic Brownian motion of a non-relativistic heavy probe quark immersed in an weakly anisotropic, strongly coupled  hot plasma.
Our computation in the bulk theory involves an explicit solution of the transverse fluctuation modes of the probe string in the low frequency regime along anisotropic
as well as isotropic directions. The above restrictions are imposed to have an analytic handle upon the computations. One might try to relax some of these restrictions, like considering general values of the parameter $a/T$. For large values of $a/T$ or, small values of $T$, the gravity background is known analytically and one might try to perform a similar computation. However, in that regime of the parameter space, the quantum fluctuations will dominate over the random fluctuations. For intermediate values of $a/T$ no analytical results are available and one will have to fall back upon numerical means right from the outset. It might also be possible that some of the  results obtained in this paper, get modified away from these limits we have considered. Hence, it might be interesting to investigate the Brownian motion in more general scenarios\footnote{In a recent paper \cite{Giataganas:2013zaa}, the authors study the relativistic Langevin diffusion of a heavy quark in strongly coupled, anisotropic Yang-Mills plasma for both small and large values of the anisotropy parameter.}. The analytic solution is obtained using matching boundary techniques. Recently it has been shown that in the presence of a background electric field in the bulk, the worldsheet of the probe open string develops an
induced horizon structure. This is also true when the probe string possesses
a non-trivial velocity profile \cite{Sonner:2012if,Kundu:2013eba}. Since we have not
considered such configurations, our solution smoothly interpolates from
the boundary to the black hole horizon.
The only horizon structure is embedded in the black hole background. 
It is important to note that if we could precisely measure the Brownian dynamics in the boundary, it would have been a very promising step towards learning the quantum 
dynamics of black hole physics. However, that requires the knowledge of non-perturbative gauge theory correlators which is beyond the scope of this paper. 
In this work, using the holographic prescription, we have computed the drag coefficient, the diffusion constant and the strength of the random force in a low
frequency as well as non-relativistic limits. The expressions for the drag coefficient and the Langevin coefficient along the anisotropic direction
clearly signify an enhancement over the corresponding isotropic counterparts. The fluctuations
along the isotropic direction also respond to the anisotropy in the bulk. As a result, in the boundary theory, 
we observe that both the drag coefficient and the coefficient of auto correlator take lower value
compared to the case of ordinary SYM plasma. We have also checked that even in the presence of anisotropy, 
the fluctuation-dissipation theorem is still valid for random variation along
both isotropic and anisotropic directions.
Moreover, we have computed the diffusion constant and reproduced the Einstein-Sutherland relation in a holographic sense.
Before closing let us also observe an interesting qualitative agreement of our result with those obtained in the case of non-commutative Yang-Mills 
(NCYM) plasma which also has an inherent anisotropy built into it. In \cite{Fischler:2012ff} the drag force, the diffusion constant and the Langevin coefficient were 
holographically computed for strongly coupled NCYM. In the case of NCYM , an unbroken $SO(2)$ symmetry is confined to the non-commutative plane whereas for 
spatially deformed anisotropic YM plasma, the unbroken $SO(2)$ symmetry lives on the isotropic plane ($x_1$-$x_2$ plane). Therefore it is reasonable to  
compare the result in the isotropic plane in the present paper with the NCYM result. 
Within the small anisotropy approximation, it is observed that in both cases, the drag force coefficient is weaker than the one computed 
in the context of ordinary YM plasma. This observation is also true for the relevant Langevin coefficient. It is important to 
check the validity of this comparison for arbitrary strength of anisotropy. However this is beyond the scope of analytic computation and is left for a future work.
\newpage
\appendix
\section{Details of the solution along anisotropic direction by matching technique} \label{sol}
In this appendix we present the details of the solution (Eq.(\ref{asymsol})) referred to in section \ref{ani}. We employ the so-called matching technique. The solution to Eq.(\ref{mastereq}) is extremely difficult to obtain analytically for any frequency. To make the problem tractable we focus only on the behavior of the solution in the low frequency domain. In this frequency domain we resort to the matching technique whereby we find the solutions in three different regimes and then match these solutions to the leading in the frequency at the interface of two domains. To be more specific, we find solutions to Eq.(\ref{mastereq}) in the following three limiting cases: (A) Near the horizon, i.e., $y \rightarrow 1$ for arbitrary frequency and then take the low frequency limit. (B) Throughout the bulk (i.e., arbitrary $y$) but for low frequency $\nu \ll 1$ and then take the near horizon limit. We match this solution with the low frequency limit of the solution obtained in (A). Finally in (C) we solve 
the equation in the asymptotic limit $(y \rightarrow \infty)$ for arbitrary $\nu$. Then taking the low frequency limit we match it with the solution of (B). Below we elucidate the details of the solutions for each regime. 
\subsection{Near horizon limit}
In this regime we solve Eq.(\ref{mastereq}) near the horizon, i.e., in the limit $y \rightarrow 1$. In the near horizon regime, Eq.(\ref{mastereq}) simplifies to,
\be 
g_{A}''(y)+\frac{1}{y-1}g_{A}'(y)+\frac{\nu^{2}}{16(y-1)^{2}}\left[1+\frac{\tilde{a}^{2}}{24}\left(5\log2-2\right) \right]g_{A}(y)=0.
\ee
This has a solution,
\be 
g_{A}(y)=A^{\text{out}}(y-1)^{\frac{i\nu}{4}\left[1-\frac{\tilde{a}^{2}}{48}(5\log2-2) \right]}+A^{\text{in}}(y-1)^{-\frac{i\nu}{4}\left[1-\frac{\tilde{a}^{2}}{48}(5\log2-2) \right]}
\ee
where the coefficients $A^{\text{out/in}}$ correspond to outgoing and ingoing modes respectively. We normalize these modes according to Eq.(\ref{nearhorizonsol}) and expand for low frequencies to obtain,
\be \label{nhl}
g_{A}^{\text{out/in}}(y)\sim 1\pm \frac{i\nu}{4}\log(y-1)\left[ 1-\frac{\tilde{a}^{2}}{48}(5\log2-2)\right] +\z{O}(\nu^{2}).
\ee

\subsection{Low frequency limit}
Next we attempt to solve Eq.(\ref{mastereq}) in the low frequency limit but for arbitrary $y$, i.e., throughout the bulk. We can perform a series expansion in powers of $\nu$ to write the solution in the generic form,
\be 
g_{B}(y)=g_{0}(y)+\nu g_{1}(y)+\nu^{2}g_{2}(\nu)+...
\ee
Inserting this ansatz in Eq.(\ref{mastereq}), setting the coefficient of each power of $\nu$ to zero and solving the resulting equations we can find $g_{0},g_{1},g_{2}$. At the zeroth order, the equation to solve is,
\be 
g_{0}''(y)+\frac{4y^{3}}{y^{4}-1}\left[ 1+\tilde{a}^{2}\Psi(y)\right]g_{0}'(y)=0
\ee
with $\Psi(y)$ being given in Eq.(\ref{mastercoeff}). The solution to the equation for general $y$ is quite complicated and
is given by,
\be \label{fullsolg0}
\begin{aligned}
g_{0}(y)&=\frac{1}{2} C_{1} (\tan^{-1}y+\tanh^{-1}y)+C_{2}+\frac{\tilde{a}^{2}}{768 (y^4-1)}C_{1}\Bigg\{-16 y+16y^3+80 y \log2   \\
& +\log\left(1+\frac{1}{y^2}\right)(-80 y-51(y^4-1) \log(1-y)  -9(y^4-1) \log(y-1)\\
&+60(y^4-1) \log(1+y))  -(y^4-1) \bigg[\log(y-1) \big[-17
 -9 \log\left(1+\frac{1}{y^2}\right)\big]\\
 &+\log(1-y)(25+102 \log y  +17 \log(y^2+1))\\
 &-8 (1+17 \log y) \log(1+y)-8 \log(y^2+1) \log(1+y)\\
&+4\log(-i+y)\big[2 i \log(1-i y)+2 \log\left(i\frac{y+1}{y-1}\right) -i \log(4 (-i+y))\big] \\
&+4\log(i+y) \big[-2 i \log(1+i y)+2 \log\left(i\frac{y+1}{1-y}\right)+i \log(4 (i+y))\big]\bigg]  \\
&-8(y^4-1)\tanh^{-1}y(15\log2+17\log(1+y^2))\\
& +8(y^4-1) \tan^{-1}y (4-15 \log2+4 \log y)\\
&  +8 (y^4-1) \bigg[2 \text{Li}_2(1-y)-i \text{Li}_2\left(\frac{1}{2}(1+i y)\right) +2\text{Li}_2(-y)-2 i \text{Li}_2(-i y)\\
&+2 i \text{Li}_2(i y)-\text{Li}_2\left(\frac{1}{2}(-1+i) (y-i)\right) +\text{Li}_2\left(\frac{1}{2}(1+i)(-i+y)\right)\\
&-\text{Li}_2\left(\frac{1}{2}(-1-i) (i+y)\right)+i \text{Li}_2\left(\frac{1}{2}(1-iy)\right)+\text{Li}_2\left(\frac{1}{2}(1-i)(i+y)\right)\bigg]\Bigg\}+\z{O}(\tilde{a}^{4})
 \end{aligned}
\ee
with $C_{1}$ and $C_{2}$ being the constants of integration and $\text{Li}_{n}(z)$ is the Polylogarithm function. Upon taking the near horizon limit it reduces to \footnote {While taking the near horizon limit we have let $y \rightarrow 1+\epsilon$ and used the following expansion
\be \nn
\text{Li}_n\left(z+(a+ib)\epsilon \right)=\text{Li}_n\left(z \right) +\epsilon \frac{a+ib}{z}\text{Li}_{n-1}\left(z \right)+\z{O}(\epsilon^{2})
\ee}
\be \label{bnh}
\begin{aligned}
 g_{0}(y)&=C_{2}+C_{1}\left[\left(\frac{1}{8}-\frac{i}{4}\right)\pi +\frac{\log2}{4}-\frac{\log(y-1)}{4}\right]\\
 &+\frac{\tilde{a}^{2}C_{1}}{2304}\left[ 84-48 \b(2)+(24-75i-\pi)\pi -90(1-2i)\pi \log2 \right. \\
 &\left. -204(\log2)^{2}+24\log2 -24\log(y-1)+204\log2 \log(y-1)\right].
\end{aligned}
\ee
Upon comparison with Eq.(\ref{nhl}) we can extract the coefficients $C_{1}$ and $C_{2}$ as,
\be 
C_{1}=0, \qquad C_{2}=1
\ee
for both outgoing and ingoing waves. Next we proceed to find $g_{1}(y)$. Now note that $g_{1}(y)$ satisfies the same equation as $g_{0}$ and so has the same solution (Eq.(\ref{fullsolg0},\ref{bnh})) albeit with different constants of integration, but now the matching has to be done with the coefficient of $\nu$ in Eq.(\ref{nhl}). Replacing $C_{1}$ and $C_{2}$ in Eq.(\ref{bnh}) with $\tilde{C}_{1}$ and $\tilde{C}_{2}$ respectively and comparing with Eq.(\ref{nhl}) we can extract the constants for both the outgoing and ingoing waves as
\be 
\begin{aligned}
\tilde{C}_{1}^{\text{out/in}}&=\mp i\left[1+\frac{\tilde{a}^{2}}{4}\log2 \right] +\z{O}(\tilde{a}^{4}), \\
\tilde{C}_{2}^{\text{out/in}}&=\pm \left( \frac{1}{4}+\frac{i}{8}\right)\pi  \pm \frac{1}{4}i\log2\mp i\frac{\tilde{a}^{2}}{2304}\left[ -84+48 \b(2)-(24-75i -\pi )\pi \right.\\
&\left.+\left(18(1-2i)\pi  +60\log2-24\right)\log2\right]+\z{O}(\tilde{a}^{4}).
\end{aligned}
\ee
The constants so evaluated can now be used in the full solution for $g_{B}(y)$ and not just in the near horizon limit (the restriction to low frequency regime still holds, though), which now reads
\be \label{fullsol}
\begin{aligned}
g_{B}(y)&=1+ \nu\left[ \frac{1}{2} \tilde{C}_{1} (\tan^{-1}y+\tanh^{-1}y)+\tilde{C}_{2}\right]+\frac{\nu \tilde{a}^{2}}{768 (y^4-1)}\tilde{C}_{1}\Bigg\{-16 y+16y^3   \\
& +80 y \log2+\log\left(1+\frac{1}{y^2}\right)(-80 y-51(y^4-1) \log(1-y)  -9(y^4-1) \log(y-1)\\
&+60(y^4-1) \log(1+y))  -(y^4-1) \bigg[\log(y-1) \big[-17
 -9 \log\left(1+\frac{1}{y^2}\right)\big]\\
 &+\log(1-y)(25+102 \log y  +17 \log(y^2+1))\\
 &-8 (1+17 \log y) \log(1+y)-8 \log(y^2+1) \log(1+y)\\
&+4\log(-i+y)\big[2 i \log(1-i y)+2 \log\left(i\frac{y+1}{y-1}\right) -i \log(4 (-i+y))\big] \\
&+4\log(i+y) \big[-2 i \log(1+i y)+2 \log\left(i\frac{y+1}{1-y}\right)+i \log(4 (i+y))\big]\bigg]  \\
&-8(y^4-1)\tanh^{-1}y(15\log2+17\log(1+y^2))\\
& +8(y^4-1) \tan^{-1}y (4-15 \log2+4 \log y)\\
&  +8 (y^4-1) \bigg[2 \text{Li}_2(1-y)-i \text{Li}_2\left(\frac{1}{2}(1+i y)\right) +2\text{Li}_2(-y)-2 i \text{Li}_2(-i y)\\
&+2 i \text{Li}_2(i y)-\text{Li}_2\left(\frac{1}{2}(-1+i) (y-i)\right) +\text{Li}_2\left(\frac{1}{2}(1+i)(-i+y)\right)\\
&-\text{Li}_2\left(\frac{1}{2}(-1-i) (i+y)\right)+i \text{Li}_2\left(\frac{1}{2}(1-iy)\right)+\text{Li}_2\left(\frac{1}{2}(1-i)(i+y)\right)\bigg]\Bigg\}+\z{O}(\tilde{a}^{4})
 \end{aligned}
\ee
Next we can take the asymptotic limit of the full solution to arrive at,
\be \label{basym}
\begin{aligned}
g_{B}^{\text{out/in}}\sim &1\mp \frac{i\nu}{8}\left(\pi -2\log2 \right) \\
& \pm \frac{i\nu\tilde{a}^{2}}{768}\left[28-16 \b(2)-20(\log2)^{2}+\pi(-8+\pi+14\log2)+8\log2  \right]+ \z{O}(\nu^{2})\\
& \mp\frac{1}{y^{3}}\left[ \frac{i\nu}{3}\left(1+\frac{\tilde{a}^{2}}{4}\log2\right)+\z{O}(\nu^{2})\right]+\z{O}(1/y^{4}).
\end{aligned}
\ee

\subsection{Asymptotic limit}
Finally, we are to solve Eq.(\ref{mastereq}) in the asymptotic limit, i.e., near the boundary where the gauge theory lives. We attempt a power series in the form,
\be 
g_{C}(y)=k_{0}+k_{1}/y+k_{2}/y^{2}+k_{3}/y^{3}.
\ee
It turns out the only the constants $k_{0}$ and $k_{3}$ are independent and the solution assumes the form,
\be 
g_{C}(y)=k_{0}\left[1+\frac{\nu^{2}}{2y^{2}}+\z{O}(1/y^{4}) \right]+k_{3}\left[\frac{1}{y^{3}}+\z{O}(1/y^{5})\right].
\ee
Matching the coefficients with Eq.(\ref{basym}) in the low frequency limit furnishes the two undetermined constants $k_{0}$ and $k_{3}$ as follows,
\be \label{casym}
\begin{aligned}
k_{0}^{\text{out/in}}= &1\mp \frac{i\nu}{8}\left(\pi -2\log2 \right) \\
& \pm \frac{i\nu\tilde{a}^{2}}{768}\left[28-16 \b(2)-20(\log2)^{2}+\pi(-8+\pi+14\log2)+8\log2  \right]+ \z{O}(\nu^{2})\\
k_{3}^{\text{out/in}}=& \mp \left[ \frac{i\nu}{3}\left(1+\frac{\tilde{a}^{2}}{4}\log2\right)+\z{O}(\nu^{2})\right].
\end{aligned}
\ee
The final result is then given in Eq.(\ref{asymsol}).
\section*{Acknowledgements}
The authors would like to acknowledge  Bala Sathiapalan, Diego Trancanelli, Shibaji Roy and Sudipta Mukherji for various fruitful discussions.

\end{document}